# Helping introductory physics students connect physics with humanities, art, social sciences, and everyday life

Brooke Rouret, Jaya Shivangani Kashyap, Jeremy Levy, and Chandralekha Singh
University of Pittsburgh, Pittsburgh, PA, 15260

**Abstract.** In this article, we reflect upon our positive experiences incorporating or working on extra credit projects in algebra-based introductory physics that asked students to connect physics with humanities, social sciences, and everyday life. We give an example of a student project that reflects their creativity and ingenuity and encourages other instructors to offer similar projects.

## Introduction

Algebra-based introductory physics courses that are geared towards bioscience and other health-related majors as well as non-science majors focus on helping students learn foundational physics concepts and develop problem solving and reasoning skills. Instructors can also use these courses as opportunities for students to consider how physics connects with humanities, social sciences, and everyday life, which can foster their creativity and critical thinking skills [1-10]. Here we give an example of a student project (who is a co-author of this paper) given in an introductory physics course that helps them weave their interests and hobbies with physics as part of an extra credit assignment. These types of projects bring out their creativity while helping them think deeply about physics. We note that it is important to give students grade incentives, e.g., extra credits, so that all students take this task seriously. For the work described here, a very small amount of the total grade was allocated for an extra-credit assignment. Students could choose their approach, if it wasn't computer-generated. Students wrote creative and inspiring stories and poems, sketched drawings, created sculptures or video shorts, among others relating physics with art, humanities and everyday life in deep ways.

## An Example of Student Work

The example is from a class that was taught before generative AI such as ChatGPT was introduced. The example of student work we showcase is a story about empathy and friendship between two friends who are different shapes and how one friend (who was shaped like a hoop) was able to cheer his friend up (who was shaped like a solid cylinder) by knowing the physics of rolling motion of different shapes. Students have already seen the in-class demonstration about how a solid cylinder will outpace a hollow one, and they have also solved physics problems using the physics equations that prove this fact. In this Art and Physics submission, the student (a co-author) created a story featuring a hoop and a solid cylinder. The hoop-shaped friend knew that if he asked his solid cylinder-shaped friend to do a race down an inclined plane, his friend would win the rolling race down the inclined plane because of their different shapes. The story accurately depicts physics, but the focus is on emotions, empathy and friendship and figuring out a way to make a friend, who is sad, feel happy using physics concepts. The story is depicted in Figs. 1 and 2.

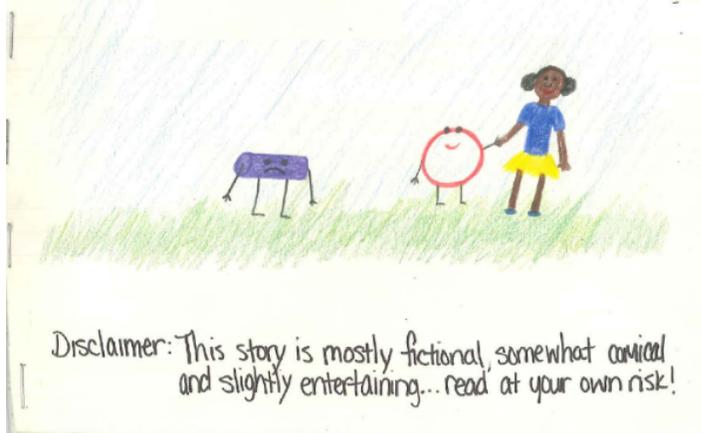

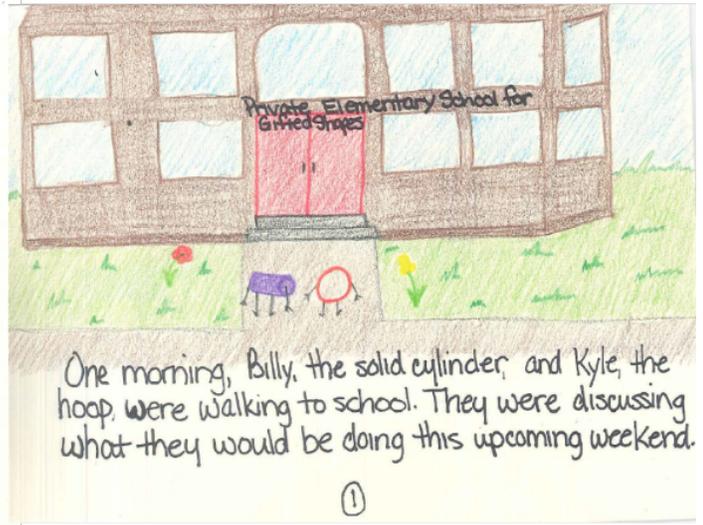

One morning, Billy, the solid cylinder, and Kyle, the hoop, were walking to school. They were discussing what they would be doing this upcoming weekend.

①

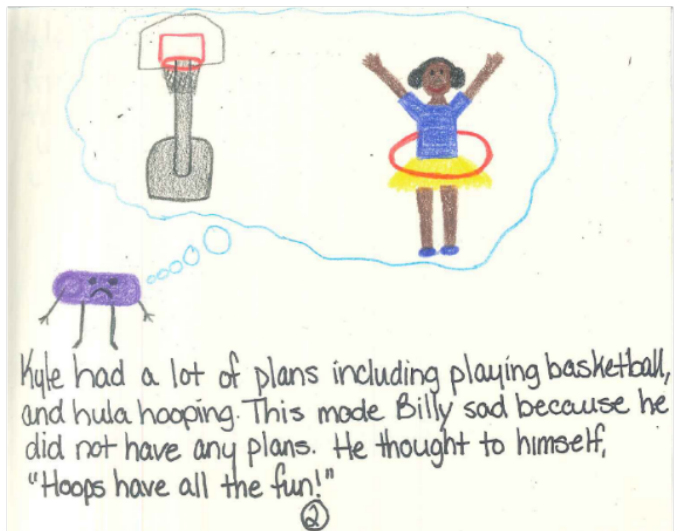

Kyle had a lot of plans including playing basketball, and hula hooping. This made Billy sad because he did not have any plans. He thought to himself, "Hoops have all the fun!"

②

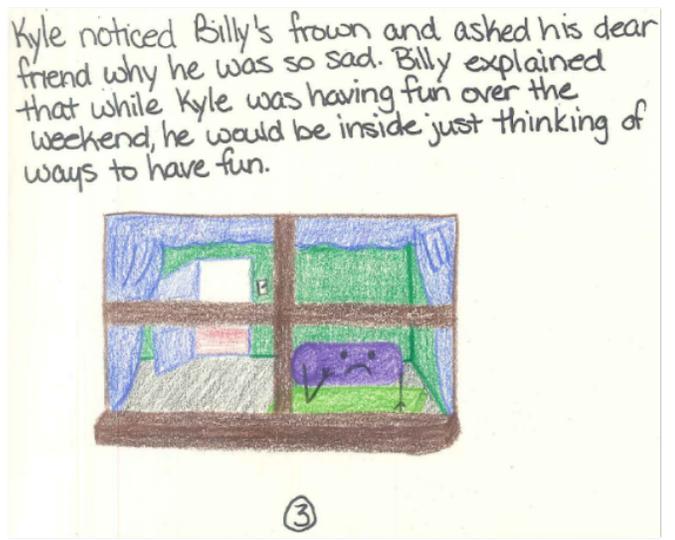

Kyle noticed Billy's frown and asked his dear friend why he was so sad. Billy explained that while Kyle was having fun over the weekend, he would be inside just thinking of ways to have fun.

③

Figure 1. The first four pages of the story

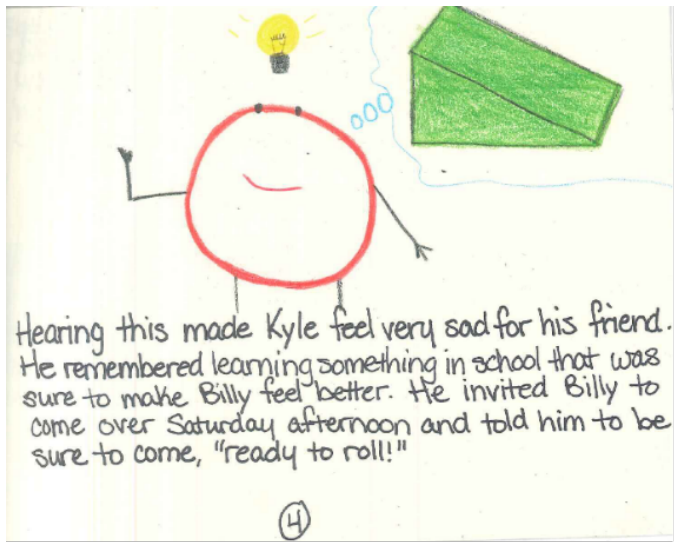 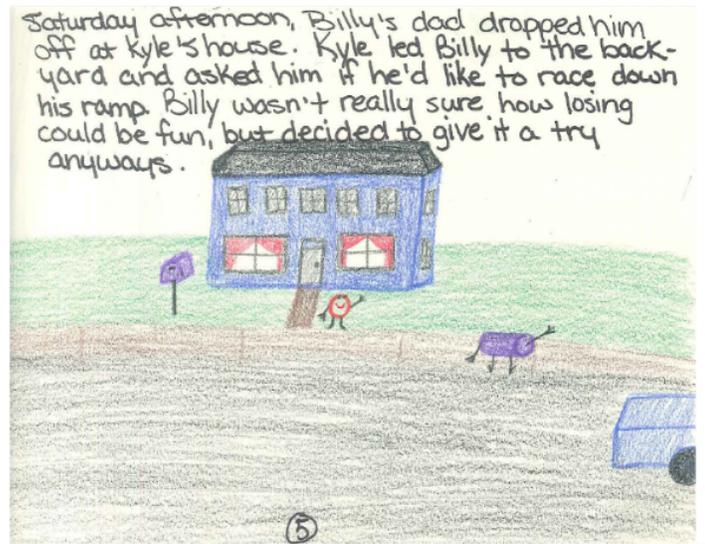
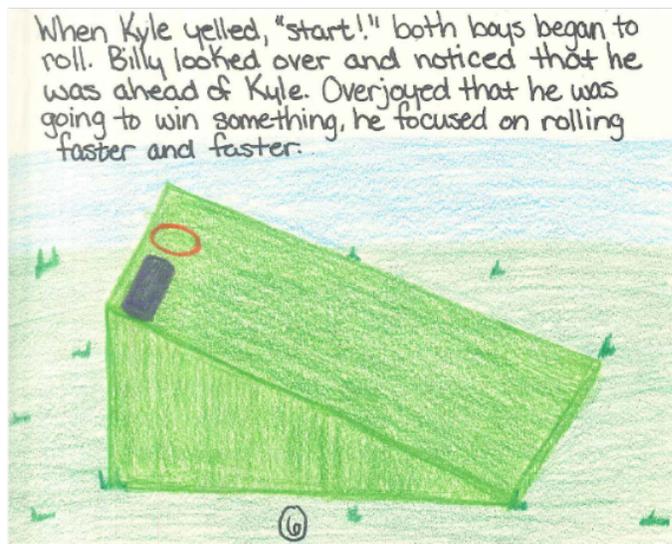 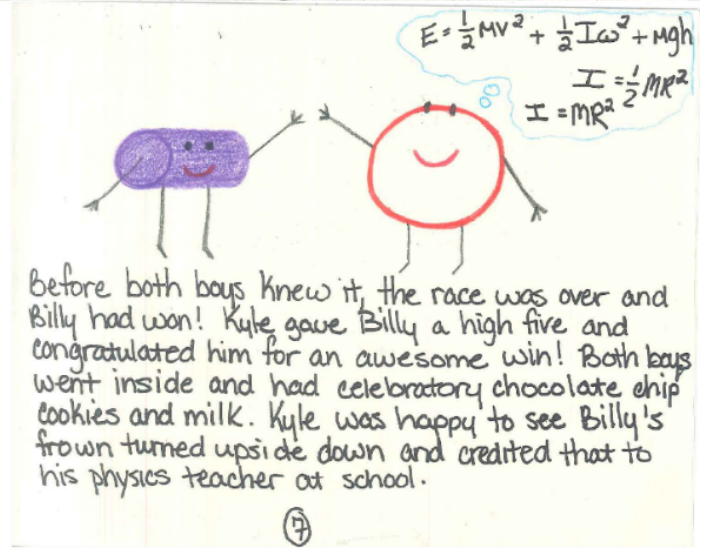

Figure 2. The last four pages of the story

**Summary and Instructional Implications**

Here we showcased an example of a student's work in an algebra-based introductory physics course that connect physics with humanities, social sciences, and everyday life. Many other students wrote poems that meaningfully incorporated physics. For example, a few lines of a very long poem after a student describes lots of physics learned in the poem are:

Making Macaroni, is as fun as making doodles.
Convection currents heat water fast, so you can add your noodles.
While blowing up a helium balloon, you think what the pressure might be.
It's quite easy to figure out, if you refer to PV=nRT.

Another student wrote a long poem titled "Physics in the World" that started with the student recollecting our classroom demonstration inviting students to cut a pizza with a knife, showing

that a pizza is very difficult to cut with a knife but easy to cut with a roller. The poem starts as follows:

"Physics is all around my life,
Like when I try to cut a pizza with a knife,
Wheels of cars undergo circular motion,
Waves have a particular frequency in the ocean."

Feedback from students were overwhelmingly positive. They suggested that students enjoyed these creative projects that made them think hard about connecting physics they were learning to art and humanities while helping them further deepen their physics knowledge during these types of contemplations. We encourage instructors to incorporate these types of extra-credit projects in their courses to help bring out student creativity while helping them connect physics and humanities and think critically about how physics concepts learned relate to everyday life.